\documentstyle[12pt,psfig]{article}
\begin{document}
\begin{titlepage}
\flushright{APLHA 99-03}
\begin{center}
\begin{huge}
\addtocounter{page}{1}
A New Technique for Sampling Multi-Modal Distributions \\
\end{huge}
\vspace{1.2cm}
K.J. Abraham \\
Department of Physics and Astronomy \\
Iowa State University \\
Ames IA 50011 \\
e-mail abraham@iastate.edu \\
\vspace{1cm}
L.M. Haines \\
Department of Statistics and Biometry \\
University of Natal \\
Private Bag X01 \\
3209 Scottsville \\
Pietermaritzburg \\
South Africa \\
e-mail haines@stat.unp.ac.za
\vspace{1.2cm}
\begin{abstract}
In this paper we demonstrate that multi-modal Probability Distribution 
Functions (PDFs) may be efficiently sampled
using an algorithm originally developed for numerical integration by
monte-carlo methods. This algorithm can be used to 
generate an input 
PDF which can be used as an independence sampler in a Metropolis-Hastings
chain to sample otherwise troublesome distributions.
Some examples in one, two, and five dimensions are worked out. 
We also comment on the possible application of our results to event generation 
in high energy physics simulations.({\em Subj. Classif.} 68U20, 65C05, 81V25,
81V15.~~{\em Keywords} Monte Carlo Optimisation, Metropolis-Hastings Chain,
Vegas Algorithm, Independence Sampler).
\end{abstract}
\end{center}
\end{titlepage}
The key to solving a wide range of optimisation problems in science 
and engineering lies
in being able to efficiently sample a (possibly very complex) PDF in one 
or more dimensions. In many cases of interest, this requires inverting 
an integral which may not be possible by 
analytical or semi-analytical means. In such circumstances, efficient 
computer algorithms are crucial. The perhaps best known such algorithm is the 
Metropolis algorithm \cite{metrop}, 
which can in principle be used to generate an accurate sample from 
any PDF no matter how complex, by a guided random walk. However, the 
Metropolis algorithm is potentially inefficent when confronted with a PDF 
with 
multiple modes, or peaks, especially if they are well seperated. As is well 
known, a very large number of random steps may needed to locate
a new mode, once one mode has been discovered, 
leading to a dramatic drop in the efficiency of the scheme. In this paper
we will show how this problem can be circumvented in a certain class of 
problems.

In order to make the subsequent discussion more clear, we will present a 
brief analysis of the weakness of the Metropolis scheme outlined in
the previous paragraph. Let
${\vec{X}}_{i}$ be some randomly choosen point in the space where the 
PDF of interest $\Pi$ (not necessarily normalised), is to be sampled. A 
new point ${\vec{X}}_{f}$ at
a distance $\delta$ from ${\vec{X}_{i}}$ is choosen and the ratio 
$\frac{\Pi({\vec{X}}_{i})}{\Pi({\vec{X}}_{f})}$ is evaluated. 
If this ratio is larger than
one, then the move $ {\vec{X}_{i}} \rightarrow {\vec{X}_{f}} $ is accepted.
Otherwise it is accepted with probability 
$\frac{\Pi({\vec{X}}_{i})}{\Pi({\vec{X}}_{f})}$. As can be imagined, locating 
a single peak of $\Pi$ can be easily accomplished. However, moving from one 
peak to another separated by a distance which is large compared with the 
stepsize $\delta$ may require a long succession of steps "against the 
grain"; the net probability of such a  sequence is sometimes so small that a 
prohibitively large number of trials may be needed in order to establish
the existence of the second peak. This in a nutshell, is the reason for 
the potential inefficiency of the Metropolis algorithm alluded to 
earlier.

One plausible remedy, varying $\delta$ with each move has been incorporated
into the Metropolis-Hastings algorithm \cite{hast}, where the sequence of 
steps is 
made on the basis of a proposal distribution. If the proposal distribution
mimics $\Pi$, then all the peaks of $\Pi$ may be found without difficulty.
However, without prior knowledge of the 
separation between the peaks of $\Pi$, it is difficult to make a
suitable choice for the proposal distribution.
In other words, $\Pi$ must be mapped out globally in the region of interest 
{\em before} it has even been studied. This requirement
may appear to present an insurmountable obstacle to the use of the 
Metropolis-Hastings algorithm; the 
rest of this paper deals with methodology we have developed to deal
with this problem.

The key to our approach is the observation that the global structure 
of $\Pi$ is required for another seemingly different problem, the 
evaluation of the definite integral of $\Pi$ over the region of 
interest. One technique for doing do which is easily adapted 
to integrands of higher dimensions is adaptive Monte Carlo simulation. 
A number of points are thrown at random along the boundaries of the 
region of interest (defining a grid) and the function is evaluated at 
these points. This process is repeated, however the second time around
the grid from the first iteration is refined so that it is finer in 
regions where the function is larger and coarser where the function is
smaller. On the third iteration, the grid 
previously obtained is further refined, and so on.
After a suitable number of iterations  a reliable estimate of the 
integral may be obtained, for a large class of integrands of 
interest. Several different
variants of this basic algorithm have been developed; we use
the VEGAS algorithm \cite{VEG}. In VEGAS the grid points are used to 
subdivide the axes into a maximum of fifty bins. The bin boundaries 
may be used to break up the region of integration into a number of 
hypercubes. Ideally, the boundaries of the hypercubes are such 
that $\Pi$ integrated over
each hypercube gives the same contribution to the definite integral 
of $\Pi$ over the region of interest. Smaller hypercubes would then correspond 
to regions where $\Pi$ is large, larger hypercubes to regions where $\Pi$
is small.

Quite apart from the definite integral, the grid information may also be 
used to define a PDF ${\cal P}$ which roughly mimics $\Pi$. 
Sampling from ${\cal P}$ is straightforward; hypercubes are picked at 
random in such a way that the probability of picking any given hypercube 
is the same for all hypercubes, and a 
random number is used to locate a point $\vec{X}$ in the hypercube by
uniform sampling. 
${\cal P}$ is defined so that it is the same for all points in a given
hypercube, and the value of ${\cal P}$ in a hypercube of volume 
$\Delta V$ is $1\over \Delta V$. More specifically, in one 
dimension a random number is used to pick a bin along the $x$ axis
in such a way that the probability of picking any bin is the same.
Then a second random number is used to pick a point within the bin,
all points within the bin sampled uniformly.
$\Delta V$ is the bin width, so ${\cal P}$ for the point chosen
is defined as the inverse of the width of the bin in which the point is 
located, independent of the precise point choosen in the bin.
In two dimensions two random numbers are used to pick an area element, 
and another two random numbers are used to pick a point in the area 
element. $\Delta V$ is now the area, so ${\cal P}$ at the point chosen
is defined to be the inverse of the area element. In effect, we have sampled
the function globally and have used VEGAS to adaptively construct  
a PDF ${\cal P}$ which is different from $\Pi$ which
nonetheless mimics $\Pi$. 
This procedure can obviously generalised to 
arbitrarily high dimensions. Regions where $\Pi$ is large (small) 
corespond the regions where $\Delta V$ is small (large) and hence to 
regions where ${\cal P}$ is large(small). 

Our strategy for sampling from $\Pi$ amounts to setting up a Metropolis-
Hastings chain using ${\cal P}$ as a proposal distribution.
From the discussion in the previous paragraph it is clear that regions where
$\Pi$ are large are more likely to be selected than where $\Pi$ is small.
A move ${\vec{X}}_{i} \rightarrow 
{\vec{X}}_{f}$ is accepted (rejected) if 
\begin{equation}
\frac{\Pi({\vec{X}}_{i})}{{\cal P}({\vec{X}}_{i})} \times
\frac{{\cal P}({\vec{X}}_{f})}{\Pi({\vec{X}}_{f})} > rn (< rn) 
\label{eq:crit} \end{equation}
where $rn$ is a random number uniformly distributed between 0 and 1.
Essentially, we are using ${\cal P}$ as an 
independence sampler for $\Pi$. This method does preserve the
condition of detailed balance and the stationary 
distribution of the resulting Markov Chain does indeed correspond to 
$\Pi$ \cite{LuT}. 
Note that the fixed step size $\delta$ plays no role whatsoever, rather 
$\delta$ varies from move to move tuned to the seperation between 
the peaks of $\Pi$. One potential objection
to this scheme is that the function must be evaluated a large number of 
times by VEGAS before a random sample can be drawn from it and it is not 
obvious whether the number of function evaluations needed is less than 
would be required in an approach with fixed step-size. This objection
will be addressed in the example we consider.

The first and simplest example we consider is a mixture of 
univariate gaussians defined in the interval $[0,22]$. The precise 
function $\Pi$ is given by \begin{equation}
.5\left\{{\cal N}(x,3,1)\right\} + .2\left\{{\cal N}(x,14,.025)\right\} + 
.3\left\{{\cal N}(x,19,.75)\right\} 
\label {eq:1dim} \end{equation}  
where $ {\cal N}(x,\overline{x},\sigma^{2})$ denotes a 
uni-variate gaussian with mean $\overline{x}$ and 
variance $\sigma^{2}$.
This function clearly has well-seperated multiple peaks; generating a 
sample from this PDF of this kind is thus liable to be problematic.

The first step in our approach is to integrate $\Pi$ with VEGAS preserving
the grid information generated by VEGAS. In this case the 
grid information is a set of 50 points in the interval $[0,22]$. The
points define bins which are such that the contribution to the definite
integral from each bin is nearly equal. As expected, the bins
are narrow (wide) where the integrand is large (small). 
$\Pi$ was evaluated $2500$ times for this purpose and a grid reflecting
the peaks in $\Pi$ was used to generate bins of varying widths. These bins
were used to define ${\cal P}$ in the interval $ [0,22]$ along the lines
just described. ${\cal P}$ thus obtained has been plotted in
Fig.~1; the correspondence between Fig.~1 and Eq.~\ref{eq:1dim} is striking.

The next step is to generate a sample from $\Pi$ using ${\cal P}$ as an
independence sampler.  The acceptance rate of the Metropolis-Hastings
chain is remarkably high, about 
80\%; {\em i.e.} about 80 \% of the moves were accepted using the 
criterion defined in Eq.~\ref{eq:crit}. This is desirable from the point
of view of minimising CPU time and reflects the accuracy with which 
${\cal P}$ mimics the underlying 
distribution $\Pi$ defined in Eq.~\ref{eq:1dim}. 
In all, $\Pi$ was evaluated a total of 15,000 times to generate a
sample. We have checked that the average value
of the random variable as well as a number of higher moments are 
correctly reproduced, within statistical error bars. This implies not only 
that all peaks 
have been discovered but crucially, that the relative weights of all the 
peaks have also been correctly reproduced. By way of comparision,
we have checked that running a Metropolis
chain with the $\Pi$ evaluated over 100 000 times with fixed step size 
does not convincingly reproduce even the first two moments. The advantage of 
our approach is clear. 

We now go on to two dimensional examples. Here a complication arises; in
dimensions larger than one the VEGAS algorithm implicitly assumes that 
${\cal P}$ is factorisable; {\em i.e.} ${\cal P}$ may be accurately
represented in the form
${\cal P} = p_{i}(x_{i})p_{j}(x_{j}) \cdots $. For
many functions of interest this is a reasonable approximation, however
if the function has a peak along a lower dimensional hypersurface other 
than a co-ordinate axis, this
approximation may be a poor one. In particular, the
VEGAS algorithm performs poorly if the function (assumed to be 
defined in a hypercube) has a peak along a diagonal of the hypercube. 
However, this does not mean that the distribution ${\cal P}$ generated from
the VEGAS grid cannot be used to sample from $\Pi$. All that happens is 
that the acceptance rate of the resulting Metropolis chain is 
lower. To illustrate this point, we consider a mixture of two bi-variate
gaussians in a square whose means lie along a diagonal.
The precise function is defined below. \begin{equation}
\Pi = 0.7\left\{{\cal G}(x,y,4,4,1,1,.8)\right\} \,+ \,
0.3\left\{{\cal G}(x,y,12,12,1,1,-.8)\right\} \label{eq:biv}\end{equation}
where $ {\cal G}(x,y,\mu_{x},\mu_{y},\sigma_{x},\sigma_{y},\rho) $ 
is defined by \begin{displaymath}
\frac{1}{2\pi\sigma_{x}\sigma_{y}\sqrt{1 -\rho^{2}}}
\exp{\frac{-1}{{2(1-\rho^{2})}}
\left[\frac{(x -\mu_{x})^{2}}{{\sigma_{x}}^{2}} +
\frac{(y -\mu_{y})^{2}}{{\sigma_{y}}^{2}} -
\frac{2\rho(x -\mu_{x})(y -\mu_{y})}{\sigma_{x}\sigma_{y}}\right] }
\end{displaymath}

The region of integration is an $(16 \times 16)$ square with one corner at 
the origin and sides along the positive $x$ and $y$ axes.
This function is not well suited to evaluation by VEGAS as both peaks lie 
along a diagonal of the square, and this is 
reflected in the fact that the acceptance rate of the Metropolis-Hastings
chain is only $\sim 23\%$. 
However, the grid
information does correctly reflect the location of both peaks and 
the values of $<x^{n}y^{m}>$ where $(m + n)\leq 6$ are corectly reproduced,
indicating 
not only that both peaks have been found, but also that the relative 
weights assigned to both peaks is correct. As a check, we have considered 
another function,
\begin{displaymath} \Pi = 0.7\left\{{\cal G}(x,y,4,4,1,1,.8)\right\}\,+\,
0.3\left\{{\cal G}(x,y,12,4,1,1,-.8)\right\} \end{displaymath}
which differs from the bi-variate gaussian in Eq.~\ref{eq:biv}
in that both peaks now lie along a line parallel to the $x$ axis. Once again, 
grid information
is used to generate a sample from which correct moments can be recovered. 
This time though, due to the more favourable location of the peaks the 
acceptance rate is almost twice as high as previously.
We see again that
an adaptive Monte Carlo approach can generate an independence sampler for 
a Metropolis-Hastings chain even when the target distribution $\Pi$ is 
two dimensional and has well seperated modes.
It is worth pointing out that modifying $\Pi$ by the introduction of 
stepping stone distributions \cite{conf} has been suggested as a means to 
facilitate sampling PDFs of this nature; in our approach no such 
modifications are necessary.

We conclude with a brief discussion of the relevance of our
methods for event generation in experimental high energy physics 
simulations, where a sample from a potentially very complicated differential
scattering cross-section dependent on more than two variables is required.
If analytic inversion is not possible
(as is often the case), another approach such
as rejection sampling is needed. This however requires an enveloping
distribution which must be somehow obtained, either by guesswork or
possibly by using the VEGAS grid information \cite{kawab}.
Alternatively the grid information may be used to construct an importance
sampler for a Metropolis-Hastings chain which can be used to generate 
events. To test this in practise,
we have considered the example of anomalous single $t$ production 
in future $\gamma \gamma$ colliders, followed by 
$t \rightarrow b \ell \nu$ evaluated in the narrow width approximation
for the $t$ and $W$ \cite{tcver}.
The five dimensional phase space has been integrated over with VEGAS
and the resulting grid was used as an importance sampler 
to generate events along the lines of the previous examples. Neglecting 
the effects of cuts, smearing and hadronisation, we obtained an acceptance 
rate of about $75 \%$, even though no attempt whatsoever was made to optimise 
the grid. In particular, our sampling did not make any use of
simplifications resulting either from the use of the narrow 
width approximation or from the $(V-A)$ structure of weak decays.
This suggests that the methods we have outlined may
be worthwhile incorporating into event generators for high energy physics, 
at least in instances when the phase space can be integrated over with VEGAS.

\noindent{\em Acknowledgments} KJA wishes to thank Prof. Krishna Athreya for 
valuable encouragement and useful discussions, and Prof.J. Hauptman
for reading a preliminary version of the manuscript.

\newpage
\thispagestyle{empty}
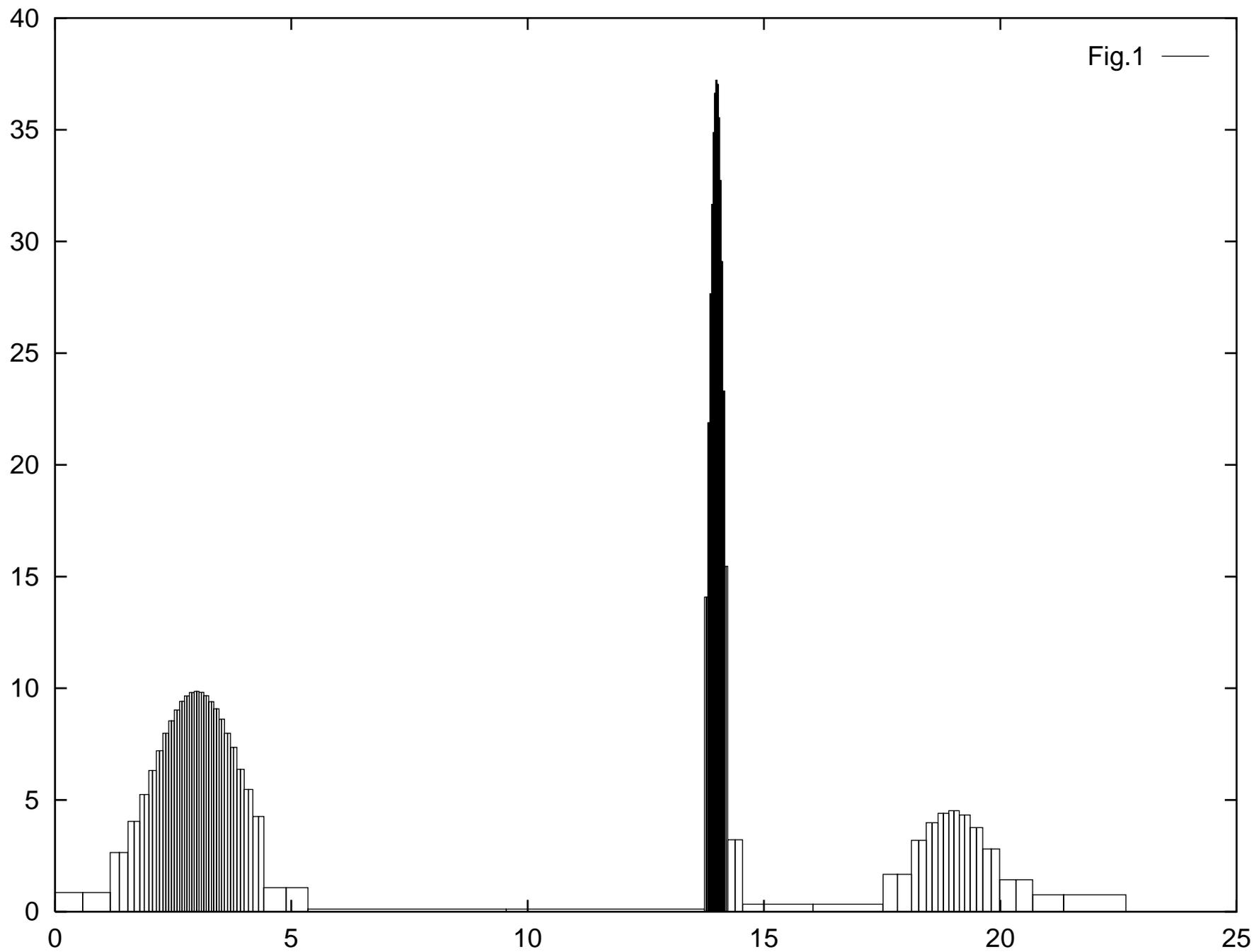
\end{document}